\documentstyle[12pt]{article}
\begin{document}
\begin{flushright}
IHEP 94-89\\
August 1994\\
Submitted to JETP Lett.\\
\end{flushright}
\begin{center}
{\bf \large $\boldmath{\bar \Lambda}$ from QCD Sum Rules for Heavy Quarkonium}
\\
\vspace*{1cm}
V.V.Kiselev\\
{\it Institute for High Energy Physics,\\
Protvino, Moscow Region, 142284, Russia,\\
E-mail: kiselev@mx.ihep.su\\
Fax: +7-095-230-23-37}
\end{center}

\begin{abstract}
Using a specific scheme of the QCD sum rules for heavy quarkonium in the
leading approximation over the inverse heavy quark mass, one gets the estimate
of the difference between the masses of the heavy meson and heavy quark
$\bar \Lambda = 0.59\pm 0.02$ GeV.
\end{abstract}

The difference between masses of the heavy meson $(Q\bar q)$ and the heavy
quark $Q$, $\bar \Lambda = m_{(Q\bar q)}- m_Q$, is the important parameter
in the Heavy Quark Effective Theory (EHQT) [1], based on the kinematical
expansion over the inverse heavy quark mass and playing an essential role
in a description of the QCD dynamics in the study of semileptonic decays
of the heavy mesons, for example, $B\to D^{(*)}l\nu$. At present, in the
framework of the QCD sum rules [2] for the heavy mesons, the $\bar \Lambda$
estimates give the value $\bar \Lambda = 0.57\pm 0.07$ GeV [3,4].

In the present paper we will show, that, because in the leading order
$\bar \Lambda$ determines the threshold of a hadronic continuum of the
two-current correlator in the consideration of the QCD sum rules for the
heavy quarkonium, one can get the estimate $\bar \Lambda = 0.59\pm 0.02$ GeV
with the much better accuracy due to  the detailed experimental data on the
spectroscopy of the charmonium $(\bar c c)$ and bottomonium $(\bar b b)$.

In the used scheme of the QCD sum rules [5] for the vector currents [2], we
introduce the number $n(q^2)$ for the $nS$-level in the heavy quarkonium
($n(M_k^2) = k$), so that the resonance contribution can be written in the form
\begin{eqnarray}
\Pi_V^{(res)}(q^2) & = & \int \frac{ds}{s-q^2}\;\sum_n f^2_{Vn} M^2_{Vn}
\delta(s-M_{Vn}^2)\;,
\nonumber \\
&=&  \int \frac{ds}{s-q^2}\; s f^2_{Vn(s)}\;
\frac{dn(s)}{ds}\;\frac{d}{dn} \sum_k \theta(n-k)\;.
\end{eqnarray}
Taking the average value of the derivative for the step-like function, we get
\begin{equation}
\Pi_V^{(res)}(q^2) = <\frac{d}{dn} \sum_k \theta(n-k)>\; \int \frac{ds}{s-q^2}
s f^2_{Vn(s)} \frac{dn(s)}{ds}\;,
\end{equation}
and, assuming, that
\begin{equation}
<\frac{d}{dn} \sum_k \theta(n-k)> \simeq 1\;,
\end{equation}
we find, that the resonance contribution, averaged by such way, is equal to
\begin{equation}
<\Pi_V^{(res)}(q^2)>  \approx  \int \frac{ds}{s-q^2}
s f^2_{Vn(s)}\; \frac{dn(s)}{ds}\;.
\end{equation}
As for the theoretical part of the QCD sum rules for the vector currents,
first, in the leading approximation over the inverse heavy quark mass,
we neglect the power corrections from the quark-gluon condensates, which
give small contributions into the leptonic constants $(\le 15\% [2])$,
and, second, we take into the account the Coulomb-like
$\alpha_S/v$-corrections, which are important in the heavy quarkonium, where
$v\to 0$, so the corrections have the form of the factor
\begin{equation}
F(v) = \frac{4 \pi}{3}\;\frac{\alpha_S}{v}\;
\frac{1}{1-\exp (-\frac{4 \pi \alpha_S}{3 v})}\;,
\end{equation}
that correctly restores the $O(\alpha_S/v)$-contribution, obtained in the
QCD perturbation theory [2].

Then near the threshold of the heavy quark production, we will have
\begin{equation}
\Im m \Pi_{P,V}^{(pert)}(s) \simeq \alpha_S\; 8\mu^2\;.
\end{equation}
where
$\mu = m_Q m_{Q'}/(m_Q+m_{Q'})$, $s\simeq M^2\simeq (m_Q+m_{Q'})^2$.
Making equal the theoretical part of the sum rules and the averaged
contribution of the resonances and assuming, that the hadronic continuum
contribution is equal to the calculated part in the QCD perturbation theory at
$\sqrt{s} \ge m_{(Q\bar q)} + m_{(\bar Q' q)} \simeq m_Q + m_{Q'} +
2 \bar \Lambda$, we find, that
\begin{equation}
\frac{f_n^2}{M_n} = \frac{\alpha_S}{\pi} \; \frac{dM_n}{dn}
\; \biggl(\frac{4\mu}{M}\biggr)^2\;.
\end{equation}
As it has been noted in the phenomenological potential models of the
heavy quarkonium [6], the quarkonium state density does not depend on the
quark flavours
\begin{equation}
\frac{dn}{dM_n} = \phi (n)\;,
\end{equation}
since the potential is close to the logarithmic one. $\alpha_S\simeq 0.2$
weakly changes under the step from the charmonium ($\alpha_S(\bar c c) \simeq
0.22$) to the bottomonium ($\alpha_S(\bar b b) \simeq 0.18$), so that
with the accuracy up to the logarithmic corrections and in the leading
order over the inverse heavy quark mass, we have the scaling relation [5,7]
\begin{equation}
\frac{f_n^2}{M_n}\; \biggl(\frac{M}{4\mu}\biggr)^2 = const.\;,
\end{equation}
that for the quarkonium with the hidden flavour, $4\mu/M=1$, converts into
the equation
\begin{equation}
\frac{f^2}{M} = const.\;,
\end{equation}
that is in a good agreement with the empirical data on the $\psi$- and
$\Upsilon$-particles.

Further, the integration of the resonant contribution by parts gives [8]
\begin{equation}
\frac{df_n}{f_n dn} = - \frac{1}{2n}\;,
\end{equation}
so that
\begin{equation}
\frac{f^2_{n_1}}{f^2_{n_2}} = \frac{n_2}{n_1}\;,
\end{equation}
that is also in a good agreements with the experimental values of the
leptonic constants for the $nS$-states, lying below the level of the hadronic
continuum.

Expressing $f_n$ through $f_1$, one can get [9]
\begin{equation}
\frac{dM_n}{dn}= \frac{1}{n}\; \frac{dM_n}{dn}(n=1)\;.
\end{equation}
so that
\begin{equation}
\frac{M_n-M_1}{M_2-M_1} = \frac{\ln {n}}{\ln {2}}\;,\;\;\;n\ge 2\;,
\end{equation}
that agrees with the spectroscopy. Moreover,
\begin{equation}
{M_2-M_1} = \frac{dM_n}{dn}(n=1)\; {\ln {2}}\;.
\end{equation}

Taking the integrals of the sum rules at $q^2=0$, in the leading approximation
we find [10]
\begin{equation}
\frac{dM_n}{dn}(n=1) \simeq \frac{2\bar \Lambda}{\ln{n_{th}}}\;,
\end{equation}
where $n_{th}$ is the number of $nS$-levels below the continuum threshold.
This number weakly depends on the quark flavours.

The $1/m_Q$-expansion must be more effective for the $b$-quarks, so,
accepting $n_{th}=4$, as in the $(\bar b b)$-system, we obtain
\begin{equation}
M(2S) - M(1S) = \bar \Lambda\;.
\end{equation}

Note, that the analogous sum rules for the pseudoscalar currents lead to
the same results, so that, neglecting the spin-dependent splitting and
using the data on the spectroscopy of the $\psi$- and $\Upsilon$-particles, we
find, that
\begin{equation}
\bar \Lambda = 0.59\pm 0.02\;\;GeV\;.
\end{equation}

Thus, the sum rules allow one to state the scaling relations and the heavy
quark flavour-independence for the density of the $S$-wave quarkonium levels,
and, on this base, to make the estimate of the $\bar \Lambda$ parameter
with the accuracy, close to the level of an uncertainty, related with the
role of both the nonperturbative power corrections and the logarithmic one
(see ref.[3]).
\newpage
\small
\centerline{{\bf References}}

\begin{enumerate}
\item
M.B.Voloshin and M.A.Shifman, Sov.J.Nucl.Phys. {\bf 45}, 292 (1987),
{\bf 47} 511 (1988);\\
N.Isgur and M.B.Wise, Phys.Lett. {\bf B232}, 113 (1989),
{\bf B237}, 527 (1990).
\item
M.A.Shifman, A.I.Vainstein, V.I.Zakharov, Nucl.Phys. {\bf B147} 385,
448 (1979);\\
L.J.Reinders, H.Rubinstein, T.Yazaki, Phys.Rep. {\bf 127} 1 (1985).
\item
I.I.Bigi, M.A.Shifman, N.G.Uraltsev, A.I.Vainshtein, Preprint
CERN-TH.7171/94 (1994).
\item
E.Bagan et al., Phys.Lett. B278, 475 (1992);\\
M.Neubert, Phys.Rev. {\bf D46}, 2212 (1992).
\item
V.V.Kiselev, Nucl.Phys. {\bf B406}, 340 (1993).
\item
E.Eichten et al., Phys.Rev. {\bf D21}, 203 (1980);\\
J.L.Richardson, Phys.Lett. {\bf 82B}, 272 (1979);\\
W.Buchm\"uller and S.-H. H.Tye, Phys.Rev. {\bf D24}, 132 (1981);\\
C.Quigg and J.L.Rosner, Phys.Lett. {\bf B71}, 153 (1977);\\
A.Martin, Phys.Lett. {\bf 93B}, 338 (1980).
\item
V.V.Kiselev, Preprint IHEP 94-63, Protvino (1994).
\item
V.V.Kiselev, Preprint IHEP 94-71, Protvino (1994).
\item
V.V.Kiselev, Preprint IHEP 94-74, Protvino (1994).
\item
V.V.Kiselev, Preprint IHEP 94-75, Protvino (1994).
\end{enumerate}

\end{document}